\documentstyle[12pt,aps]{revtex}%
\begin{document}
\draft
\title{ISOSPIN MASS SPLITTINGS OF BARYONS\\
IN POTENTIAL MODELS}
\author{K{\'a}lm{\'a}n Varga}%
\address{Institute for Nuclear Research of the Hungarian Academy of %
Sciences\\
Debrecen, PO Box 51, Hungary\\
and RIKEN, Hirosawa 2-1, Wako, Saitama 35101, Japan} %
\author{Marco Genovese\thanks{Supported by the EU Program ERBFMBICT
950427}, Jean-Marc Richard and Bernard Silvestre-Brac}
\address{Institut des Sciences Nucl\'eaires\\
Universit\'e Joseph Fourier--CNRS-IN2P3\\
53, avenue des Martyrs, F--38026  Grenoble}\date{\today}%
\preprint{ISN--97-xxx}%
\maketitle
\begin{abstract}
We discuss the isospin-breaking mass differences among baryons, with
particular attention in the charm sector to the  
$\Sigma_c^{+}-\Sigma_c^0$,
$\Sigma_c^{++}-\Sigma_c^0$, and $\Xi_c^+-\Xi_c^0$ splittings.  Simple
potential models cannot accommodate the trend of the available data  
on charmed baryons.  More precise measurements would offer the possibility
of testing how well potential models describe the
non-perturbative limit of QCD.
\end{abstract}
%

\section{INTRODUCTION }

A successful phenomenology of the hadron spectrum has been obtained
using non-relativistic potential models, which tentatively simulate
the low-energy limit of QCD \cite{Pot,Bhad,SBS}.
The interquark potential usually contains a linear part which
describes QCD confinement and is supplemented by a Coulomb term which   
may be
attributed to one-gluon exchange.   Spin--spin,
spin--orbit and tensor terms are added, analogous to the Fermi--Breit
components of QED potential, derived from $v/c$ expansion.

There are obvious difficulties.  Large relativistic
corrections can be anticipated for light quarks.  The naive
superposition of a Coulomb and a linear term may be too schematic.
Results rely on the potential at intermediate distances ($0.1 \,
\text{fm} \lesssim r \lesssim 0.5 \, \text{fm}$) where neither the
perturbative nor the string limit holds.  Nevertheless, the success  
of
potential models indicates that somehow delicate relativistic and
field-theory effects are hidden in the parameters.  As none of the
more ambitious approaches, for instance lattice calculation
\cite{lat}, is yet able to produce very precise results, it is still
justified to use potential models as tools to analyze hadron
properties, with the hope of better understanding the  
non-perturbative
limit of QCD.

Among the observables of interest, isospin-violating mass differences
have retained much attention.  Earlier studies of these mass
differences \cite{MiY} have been reconsidered within  
constituent-quark
models.  In general, the $n-p$, $\Sigma^{-} - \Sigma^{0}$, $\Sigma^{-} -
\Sigma^{+}$, $\Xi^{-} - \Xi^{0}$ splittings of the nucleon, $\Sigma$
and $\Xi$ multiplets are well reproduced, this fixing the quark-mass
difference $\Delta m = m_d-m_u$.  Predictions for charmed baryons can then be
supplied.  Some results concerning the $\Sigma_c$ and $\Xi_c$
multiplets are shown in Table \ref{Tab1}, together with experimental  
data
\cite{PDB}.  The estimates of Wright
\cite{Wright} and of Deshpande {\sl et al.} \cite{Desh} are not  
really
potential models, they are shown only for information and comparison
with the others \cite{Itoh,Ono,Chan,LW,Don,Kalman,capstik,Isgur,JM}

Some of the models include only a fraction of the possible  
contributions,
for instance electrostatic interaction is accounted for, but the mass
dependence of the chromomagnetic interaction is neglected when  
replacing
a $d$ quark by a $u$ one. This is hardly justified.
As underlined, e.g., by Isgur \cite{Isgur},  these isospin splittings
arise from several canceling contributions, so that each effect  
should be
carefully computed and even small terms should be incorporated.

The most striking feature of Table \ref{Tab1} is the wide spread of
predictions.  Next come the observation that none of the models is
compatible with the presently available data \cite{PDB}.  In  
particular the
predicted $\Xi_c^{0} - \Xi_c^{+}$ splitting tends to be smaller than
the PDG average $6.3 \pm 2.1$ MeV \cite{PDB}.  However, a preliminary
result by CLEO indicates a smaller value $\Xi_c^{0} -
\Xi_c^{+}=2.5\pm1.7\pm1.1\;$MeV \cite{CLEO}.  This collaboration has
also detected candidates for the internal spin excitations  
${\Xi'}_{c}^{0}$ and
${\Xi'}_{c}^{+}$ (total spin 1/2 with the light-quark pair $sd$ or  
$su$ mostly
in a
spin-triplet state). Their measurements indicate a splitting
${\Xi'}_{c}^{0}-{\Xi'}_{c}^{+}\simeq 1.7\;$MeV with a large error  
bar.
There are also data from CLEO on the $\Xi_c^*$  states \cite{CLEO}.

The $\Sigma_{c}$ multiplet is the most puzzling.  The
$\Sigma_{c}^{++}-\Sigma_{c}^{0}$ splitting is usually larger than
$\Sigma_{c}^{+}-\Sigma_{c}^{0}$, while data seemingly favour the
reverse.  In other words, most models predict an ordering of
$\Sigma_c^{++}$, $\Sigma_c^{+}$ and $\Sigma_c^{0}$ which is not seen,
to the extent one can draw any conclusion from the data.

The present investigation is motivated by the discrepancy between data and  
models.  We wish to understand whether this problem 
points out a general limitation of potential models, in particular those
based on one gluon--exchange, or can be solved by reconsidering the choice of parameters and
removing unjustified approximations in the three-body problem.
 
For this purpose, first we carefully estimate the role of each
contribution to the splittings within  specific potential models.
This should measure to which extent previous calculations suffer from
neglecting some effects or treating them approximately. Then we  
analyse the sensitivity to the choice of potential, to see whether fitting the  
data can be achieved by an appropriate tuning of parameters or it is out of  
reach of this approach.  Predictions are listed for a number of isospin multiplets,
making possible a comparison with all available experimental data, and with those
one could expect to be measured in the near future.

\section{ MODEL CALCULATION}
 
 A representative quark model is proposed in Ref.~\cite{Bhad}, where the potential
  is tentatively designed to fit  both
meson and baryon spectra, using the empirical rule
\begin{equation}
V_{QQQ}(\vec{\rm r}_1,\vec{\rm r}_2,\vec{\rm r}_3)
={1\over2} \sum_{i<j}V(\vert \vec{\rm r}_i-\vec{\rm r}_j\vert).
\label{one-half-rule}
\end{equation}
In \cite{Bhad}, the quark--antiquark potential reads
\begin{equation}
V(r)=-{\kappa\over r}+\lambda r^p-\Lambda+{2\pi\kappa'\over 3  
m_1m_2}\,
\tilde{\delta}(r,r_0)\,\vec{\sigma}_1\cdot\vec{\sigma}_2,
\label{potential-formula}
\end{equation}
with $\tilde{\delta}(r,r_0) = \exp(-r/r_0) / (4\pi r_0^2 r)$ being
a smeared form of the contact term. The parameters are (in the units used   
by the authors \cite{Bhad}): 

\begin{eqnarray}
\label{BCN}
&p=1,\quad \kappa=102.67\,{\rm MeV\, . fm} ,\quad \kappa ' =6\times  
102.67\,{\rm
MeV\, .fm}  ,\nonumber\\
{\rm (BCN)}\hspace{.5cm}&\quad \lambda =1/(0.0326)^2  \,{\rm  
MeV/fm},\quad
\Lambda = 913.5\, {\rm MeV},\quad r_0^{-1}=2.2\,{\rm fm}^{-1},\\
& m_q= 337  ,\quad m_s=  600  ,\quad m_c= 1870  , \quad m_b=5259\,{\rm  
MeV }
\nonumber.
\end{eqnarray}

One of the difficulties in the above model is that the spatial extension
$r_0$ of the spin--spin term is too large  
to
describe a short-range interaction between heavy quarks. As a  
consequence,
the $J/\Psi-\eta_c$ hyperfine splitting is not well reproduced.
This is why Ref.~\cite{SBS}, following for instance  
Ref.~\cite{Ono82},
introduces a flavour dependence in $r_0$, namely
\begin{equation}
r_0(m_i,m_j)= A \left (  {2 m_i m_j \over m_i + m_j } \right )^{-B} \, ,
\label{r0}
\end{equation}
while  a Gaussian form
$\tilde{\delta}(r,r_0)=\exp(-r^2/r_0^2)/(\pi^{3/2}r_0^3)$ is now  
adopted.
Both models AL1 and AP1 of Ref.~\cite{SBS} fit very well the meson  
spectrum.
The parameters are
\begin{eqnarray}
\label{AL1}
&p=1,\quad \kappa=0.5069,\quad \kappa ' = 1.8609 \quad \lambda =  
0.1653\,{\rm
GeV}^2,\nonumber\\
{\rm (AL1)}\hspace{.5cm}& \Lambda = 0.8321\, {\rm GeV},\quad  
B=0.2204,\quad
A=1.6553\,{\rm GeV}^{B-1},\\
& m_q=0.315   ,\quad m_s=0.577    ,\quad m_c=1.836 ,    \quad  
m_b=5.227\,   {\rm
GeV}\nonumber ,
\end{eqnarray}
and
\begin{eqnarray}
\label{AP1}
&p=2/3,\quad \kappa=0.4242,\quad \kappa ' = 1.8025,\quad \lambda =  
0.3898\,{\rm
GeV}^{5/3},\nonumber\\
{\rm (AP1)}\hspace{.5cm}& \Lambda = 1.1313\, {\rm GeV},\quad  
B=0.3263,\quad
A=1.5296\,{\rm GeV}^{B-1},\\
& m_q=0.277   ,\quad m_s=0.553    ,\quad m_c=1.819  ,  \quad  
m_b=5.206\,   {\rm
GeV}\nonumber.
\end{eqnarray}
None of these models include tensor forces,  since
this interaction is not expected to give important contributions in hadron
spectroscopy, at least for ground states \cite{IKK}.

For studying isospin breaking, we have allowed for $m_u \neq m_d$
and added to the potential the electrostatic interaction between  
quarks. The difference $\Delta m = m_d - m_u$ between $d$ and $u$  
quark masses
has been adjusted to reproduce neutron--proton and $\Sigma^{-} -  
\Sigma^{+}$ mass splittings. In particular, we have taken
$m_u= 327$ MeV and $m_d= 338$ MeV for AL1, 
$m_u= 337$ MeV and $m_d= 353.85$ MeV for BCN
and $m_u= 277$ MeV and $m_d= 300.5$ MeV for AP1. 
 Our estimate of  $\Delta m$ for constituent 
quarks is larger than the common wisdom for current quarks, $\Delta m 
\simeq 4$ MeV.
However, it must be noticed that dressing quarks modifies this quantity,
for the cloud of virtual states depends on the flavour of the quark
it is surrounding \cite{Fey,BBFG}.

Baryon masses have been obtained using two reliable numerical methods
\cite{Varga,SBp}. In the first case every contribution was included non-perturbatively
in the variational procedure, while in the second one the electromagnetic terms were
treated perturbatively.
The perfect agreement of two results indicates that  
a good
convergence has been reached. 

 The various contributions to $n-p$ and to
charmed baryons mass differences are shown in Table \ref{Tab2}, for the
specific model AL1. As
hinted previously delicate cancellations occur, requiring an accurate
treatment of each term. In a flavour-independent potential the energy 
of a given state decreases when any of the constituent masses is 
increased, i.e. $(\Delta T + \Delta W + \Delta B ) / \Delta m < 0$ in 
the notations of Table \ref{Tab2}. This is observed in our calculations,
 though in the $\Xi_c$ case, the flavour dependence of 
spin--spin term goes in the opposite direction. As for the kinetic 
energy itself, one expects \cite{Rosner} $\Delta T / \Delta m >0$
if confinement dominates the binding process, and $\Delta T / \Delta m 
<0$ when the Coulomb part becomes more influential.

Before discussing the results obtained for baryons, we have  
investigated the splittings among mesons, for instance $D^+(c\bar{d})-D^0(c\bar{u})$.
The results are shown in Table~\ref{Tab-Mes}, for the three models  
BCN, AL1 and AP1. $\Delta m$ is adopted to reproduce $m_n - m_p$ in the
baryon sector. 
 An acceptable agreement is found for AL1, while BCN and AP1 seem 
to be disfavoured, leading to a larger overestimation
of the $K^0 - K^+$ splitting.
Actually, AL1 is somehow an improvement of BCN, while 
the confining part $\propto r^{2/3}$ of AP1 does not agree with lattice
 results and fails for heavy mesons as well. 
In the following AL1 will be our benchmark, the results for AP1 and BCN 
will also be given in order to show how stable are the predictions 
respect to the choice of the potential.

Our  results for baryons splittings are shown in Table 
\ref{Tab3}.

A check of the consistency of our calculation is provided by
the relations listed by Franklin \cite{Franklin}. When one switches off
the electromagnetic
interaction, one deals with energies which evolve continuously when going,
for instance, from ($Quu$) to ($Qdd$) via ($Qud$). Hence a mass combination

\begin{equation}
\delta({\Sigma_Q})= (Quu)+(Qdd)- 2 (Qud)
\label{franklin}
\end{equation}
 receives contributions mostly from Coulomb effects.
 If the latter are treated at first order with a wave function ($Qqq$) 
properly averaging that of ($Quu$), ($Qdd$) and ($Qdu$), then 
$\delta({\Sigma_Q}) \simeq \alpha \langle r^{-1}_{qq} \rangle $, i.e.
the charge of heavy quark $Q$ disappears.

If furthermore the ($qq$) part of the of the ($Qqq$) wave function does
 not depend much on the mass of the heavy quark $Q$ and on the 
coupling of the spin of $Q$ to the ($qq$) spin triplet, then 
$\delta({\Sigma_Q})$ should be approximately the same for
 $\Sigma$, $ \Sigma ^*$, $\Sigma_c$ or $\Sigma_b$ multiplets. 
This is again rather well  verified in our calculation.
 
The possible dependence of the ($qq$) distribution on the mass of the 
third quark $Q$ is discussed by Rosner \cite{Rosner} as a ``three-body effect".
It was investigated previously \cite{Cohen,RT} in the literature, not for
$\langle r^{-1}_{qq} \rangle $, but for the somewhat similar matrix element 
$\langle \delta^{(3)} ( \vec{ \rm r}_{qq}) \rangle $  that enters the calculation of the 
hyperfine splittings in usual quark models. In an approach \`a la 
Breit--Fermi,
the ratio

\begin{equation}
R= {2 \Sigma^* + \Sigma - 3 \Lambda \over 2 \Delta - 2 N}
\label{R}
 \end{equation}
reveals  the ratio of the ($qq$) short-range correlations in ($sqq$) 
and ($qqq$). Similarly the ratio
\begin{equation}
R '= {\Xi^{*} - \Xi \over \Sigma^* -\Sigma}
\label{R2}
\end{equation}
gives a comparison of ($qs$) correlations in ($ssq$) and ($sqq$).
The experimental values $R \simeq 1.04$ and $R' \simeq 1.12$, as well as the 
detailed three-body calculations \cite{Cohen,RT} show that, as conjectured by Rosner 
\cite{Rosner}, the $q_1$ and $q_2$ quarks tend to bind more intimately
within ($q_1 q_2 q_3$) when 
$q_3$ becomes heavier. The effect is about 5 to 10 $\%$ when $q_3$ changes from 
ordinary to strange, may be slightly more from strange to charmed.
In would be desirable to reach a deeper understanding of this property, 
beyond numerical investigations.
A possible starting point is given by the harmonic oscillator 

\begin{equation}
H = p_1^2 + p_2^2 + \alpha p_3^2 + 
r_{12}^2 + \beta ( r_{13}^2 + r_{23}^2) \, ,
\label{HHO}
\end{equation}
where the $r_{12}$ dependence of the wave function  factorizes out
 and is easily shown to be independent of the inverse mass $\alpha$ (but 
it does depend on the strength $\beta$).

\section{DISCUSSION}

The isospin-violating splittings of light and heavy baryons are shown 
in Table \ref{Tab3}. The $\Sigma^{-}- \Sigma^{0}$ and
$\Delta^{0} - \Delta^{++}$ splittings and even
(within large errors) the electromagnetic splittings
for the excited states $\Sigma^*(1385)$
and $\Xi^*(1530)$ come out in good
agreement with the experimental data \cite{PDB}. Some problems appear
however for charmed baryons.

More precisely, while the experimental datum $\Sigma_c^{++} - \Sigma_c^{0}=
0.8 \pm 0.4$ MeV is well reproduced, one finds an, albeit small (of  
the
order of $ -0.5$ MeV), negative $\Sigma_c^{+} - \Sigma_c^{0}$,  
at
variance with the experimental datum $\Sigma_c^{+} - \Sigma_c^{0} =  
1.4
\pm 0.6$ MeV. Also the result for $\Xi_c^{+} - \Xi_c^{0} = 2.2$ MeV
is smaller than the, still rather imprecise, experimental datum
$\Xi_c^{+} - \Xi_c^{0} = 6.3 \pm 2.3 $ or $4.7 \pm 2.1$ MeV, where  
the first
number corresponds to the particle data group average and the second  
to their fit. The problem is not solved using AP1 or BCN.

To summarize at this stage, the splittings of charmed baryons do not agree
with experimental results, when they are calculated from potential models
supplemented by electrostatic forces and a mass difference $\Delta m$
between $d$ and $u$ quarks.

Reasonable changes of light quark masses do not modify substantially
this situation.

An effect which we have neglected up to now is the  
electromagnetic
dipole--dipole interaction between quarks, whose dominant term
(neglecting the small contribution of components
of the wave function with non-vanishing angular momentum) is:
\begin{equation}
-{2 \pi \over 3} {q_i q_j \alpha \over m_i m_j}
\delta(r_{ij}) \vec{\sigma}_i \cdot \vec{\sigma}_j \, 
\label{eq:dip-dip}
\end{equation}
where $q_i$ are quark charges in units of electron charge.
Albeit surely present, one
expects it to be smaller than usual Coulomb interaction.
Our numerical results including this dipole--dipole term are shown in 
\ref{Tab3}. We used a regularized form $\tilde{\delta}$  
for $\delta(r)$, taken to be the same as
 for the strong  spin--spin force.
The magnetic contribution goes in the right  
direction, but remains too small to push the computed masses 
significantly closer
to the experimental ones. 
 
This problem raises the question whether some contribution has been
forgotten.
For example, in some models adjusted to reproduce meson and baryon  
masses simultaneously
introduce in the baryon sector an {\sl ad-hoc} 3-body term of the  
form
\cite{Bhad,SBS}
\begin{equation}
D_3 + { A_3 \over (m_1 m_2 m_3)^{b_3}} \, .
\label{eq:threebody}
\end{equation}
This parameterization is purely empirical. For the AL1 model,
the parameters are $D_3=0.07376$,
$A_3=-0.05546$ and $b_3=1/4$.

As it depends on masses, this term gives a contribution
to isospin breaking effects as well.
The 3-body term (\ref{eq:threebody}) slightly improves the  
description
of the electromagnetic splittings of light baryons. However,
how is evident by inspecting Eq. (\ref{eq:threebody}), the  
contribution
$\Sigma_c^{+} - \Sigma_c^{0}$ goes in the wrong direction.
When the 3-body term (\ref{eq:threebody}) is accounted for one obtains
typically
$\Sigma_c^{+} - \Sigma_c^{0} \simeq -0.7$ MeV (instead of $-0.55$ MeV), and
$\Xi_c^{0}-\Xi_c^{+} \simeq 1.5$ MeV (instead of $2.58$ MeV),
with little dependence on the choice of parameters.
Of course, one could think of more complicated three-body  
interactions,
but their form remains completely arbitrary and somehow the 
appealing features of potential models are lost once one violates
flavour independence and gives up the link between quark--quark and
quark--antiquark forces.

Another possibility which can be explored is the running of
$\alpha_s$, which leads to a reduced coupling when heavy quarks  
appears
for the scale is chosen to be proportional to the masses involved
(of course problems related to the precise choice of the scale and to
the unknown $\alpha_s$ behaviour at small scales emerge).
Such an effect would decrease the strenght of the spin-spin
term involving heavy quarks, but this would not go in the right 
direction for changing the order of $\Sigma_c$ states.

Finally, we discuss now two interaction terms which have been
contemplated in addition or in replacement of chromomagnetism.

The first comes from the contribution of instantons.
The non-relativistic form of a
potential mimicking 't Hooft interaction \cite{tHooft} has been
elaborated in Ref.s \cite{Dorokhov,Bonn}. The value of the coupling
must however be fixed phenomenologically.
Interesting results have been obtained on hadron spectroscopy
with models including this instanton term replacing
\cite{Dorokhov,Bonn} or supplementing \cite{SBS2} the chromomagnetic  
force.

However, when the instantonic potential is considered only  
as a further correction to Eq. (\ref{potential-formula}), it
does not contribute
substantially to $\Sigma_{c}$ mass splittings,
for it is inversely proportional to the quark masses
and vanishes for a quark pair with spin $1$. Thus
it cannot help solving the
problem of $\Sigma_c$ splittings. It gives, anyway, a positive contribution, albeit
not to be expected quite large (of course the precise numerical value
will depend on the choice of the coupling), to $\Xi_c^{+} -  
\Xi_c^{0}$.

The second type of interaction deals with meson exchange between  
quarks.
There is a rich literature on the subject, which has recently be  
 revisited
by Glozman and collaborators (see e.g. Refs. \cite{Gloz} and  
references
therein), who have adopted rather an extreme point of view where the
chromomagnetic force is completely removed.
These authors obtain a surprisingly good fit to light and strange  
baryons.
In this approach, the study of electromagnetic splittings of baryons
is somewhat reminiscent of isospin violating effects in nuclear  
physics,
where one accounts for a difference between $\pi^{\pm}$ and $\pi^0$
masses  and their couplings to nucleons.
This remains to be studied. However, the extension of Glozman model  
to heavy
baryons seems problematic, notwithstanding some initial attempts
\cite{Gloz2}.

\section{OUTLOOK}

In conclusion, we find that non-relativistic potential models  
do not permit to reproduce the data on $\Sigma_c^{+} -
\Sigma_c^{0}$ and $\Xi_c^{+} - \Xi_c^{0} $
mass splittings,
despite the good agreement obtained for light baryons.
Of course experimental data need further confirmation, 
and need to be extended to beauty and double-charm sectors,
to see if this discrepancy persists.

Previously, Franklin \cite{Franklin} pointed out that the experimental
data on charmed baryons violate mass relations which are expected to 
hold within large class of quark models.
We have checked that the mass obtained from an accurate solution of
the three-body problem fulfill the Franklin relations. So if the 
difficulty persists, its solution should be searched in a intrinsic
 limitation of usual quark models, for instance in the need for new 
dynamical contributions, such as electromagnetic 
penguins \cite{Penguins}. 

The present situation is somewhat a paradox. The $\Lambda_c$ and 
$\Omega_c$ and the average $\Sigma_c$, $\Xi_c$ states are reasonably 
described by simple potentials, i.e., one seemingly
controls the behaviour of the 
ground-state baryons when an ordinary quark is replaced by a charm one.
Meanwhile one does not understand the effect of a more modest move,
 when a up quark is changed in a down one.

\subsection*{ Acknowledgements} Stimulating discussions with  
Fl.~Stancu and S.~Pepin are gratefully acknowledged, as well as  
correspondence from S.C.~Timm, J.~Franklin and T.~Goldman.

\listoftables
\begin{table}
\caption{\label{Tab1} Predictions of different  models for charmed
baryons
electromagnetic mass splittings}
\begin{tabular}{lccc}
 Model & $\Sigma_c^{++} - \Sigma_c^{0}$& $\Sigma_c^{+} -
\Sigma_c^{0}$ & $\Xi_c^{0} - \Xi_c^{+}$\\
\hline
Experiment \protect\cite{PDB} & $0.8 \pm 0.4\;$MeV & $1.4 \pm
0.6\;$MeV& $ 6.3
\pm 2.1\;$MeV\\
 Wright \protect\cite{Wright} & $-1.4 $&$ - 2.0$& 3.1 \\
Deshpande et al. \protect\cite{Desh}&$ - (3-18)$ & $- (2.5-10)$&
$4.5-12$\\
Itoh \protect\cite{Itoh} & 6.5 & 2.4 & 2.5 \\
Ono \protect\cite{Ono} & 6.1 & 2.2 & 1.8\\
Lane and Weinberg \protect\cite{LW} & $-6$ & $-4$ & 4 \\
Chan \protect\cite{Chan} & 0.4 & $-0.7$ & 3.2 \\
Lichtenberg \protect\cite{Don} & 3.4 & 0.8 & 1.1 \\
Kalman and Jakimow \protect\cite{Kalman} & $-2.7$& $-2.2$& 3.6 \\
Capstick \protect\cite{capstik} & 1.4 & $-0.2$ &  \\
Isgur \protect\cite{Isgur} & $-2$& $-1.8$& \\
Richard and Taxil \protect\cite{JM} \begin{tabular}{c}  I  \\  II
\end{tabular}&
      \begin{tabular}{c}  3  \\  $-2$  \end{tabular}&
      \begin{tabular}{c}  1  \\  $-1$  \end{tabular}&
      \begin{tabular}{c}  0  \\     2   \end{tabular}\\
\end{tabular}
\end{table}

\vskip 1cm
\begin{table}
\caption{\label{Tab2}
Different contributions (in MeV) to the neutron to proton and charmed
baryons mass differences: $\Delta m$ is the change of constituent  
quark
masses, $\Delta T$ the difference of kinetic energies, $\Delta W$ the
variation of the expectation value of the Wigner term (independent on
spin and isospin), $\Delta B$ the difference of Bartlett components
($\propto \vec{\sigma_i} \vec{ \sigma_j}$) and finally $\Delta C$ comes from the
Coulomb electric interaction. Here we use the model AL1.}

\begin{tabular}{cdddddd}
Baryons & $\Delta m$ &  $\Delta T$ & $\Delta W$ & $\Delta B$
&$\Delta C$ & Total\\
$n-p$ &\phantom{--}11 & 0.33 & --6.45 & --2.86 & --0.76 & 1.24\\
$\Sigma_c^{+}-\Sigma_c^0$& --11 & 0.56 & 7.55 & 1.39 & 1.14& --0.35\\
$\Sigma_c^{++}-\Sigma_c^0$& --22 & 0.79 & 15.78 & 2.07 & 4.53 &  
1.20\\
$\Xi_c^{0}-\Xi_c^+$&\phantom{--}11 & --3.31 &  --5.56 & 1.72 & --1.01  
& 2.83 \\
\end{tabular}
\end{table}

\begin{table}
\caption{
\label{Tab-Mes}
Comparison of the isospin-breaking splittings of mesons (in MeV)  
obtained from
several potential models: Bhaduri et al.\ (BCN); Silvestre-Brac and  
Semay (AL1
with linear confinement, AP1 with a $r^{2/3}$ confinement).}
\begin{tabular}{ccddd}
Splitting & Exp.\ (Ref.\ \protect\cite{PDB}) & BCN & AL1 & AP1 \\
\hline
$K^0-K^+$ &$3.995\pm0.034$ & 13.15& 6.64 & 9.56\\
$K^{*0}-K^{*+}$ & $6.7\pm1.2$& 1.55& 1.36 & 1.28\\
$D^+-D^0$ & $4.78\pm 0.10$ & 5.37 & 3.78 & --0.33 \\
$D^{*+}-D^{*0}$ & $2.6\pm 1.8$ & 2.44 & 2.74 & --0.16 \\
$B^0-B^-$ & $0.35\pm 0.29$ \footnote{Notice that on four available 
measurements, two are negative \cite{PDB}} & --1.46& --1.29 &--6.06 \\
$B^{*0}-B^{*-}$ & & --2.04 & --1.23 &--5.26 \\
\end{tabular}
\end{table}

\begin{table}
\caption{
\label{Tab3}
Comparison of the isospin-breaking splittings (in MeV) obtained from
several potential models: Bhaduri et al.\ (BCN); Silvestre-Brac and  
Semay (AL1
with linear confinement, AP1 with a $r^{2/3}$ confinement); Richard
and Taxil where the hyperfine interaction is treated perturbatively
(RT II has a linear central potential, RT I a $r^{0.1}$ one).  In the
column AL1 + dd, the magnetic dipole--dipole interaction between
quarks  is accounted for, in addition to the electrostatic potential.  
Some
redundant splittings are shown for the ease of discussion.}
\begin{tabular}{ccdddddd}
Splitting & Exp. [Ref] & BCN & AL1 & AP1 & RT II & RT I & AL1 + dd \\
\hline
$n-p$ &
\begin{tabular}{c}1.293318$\pm$\\ 0.0000009\protect\cite{PDB}\\  
\end{tabular}
&
1.38 & 1.16 & 1.29 & 1.2 & 1.3 & 1.24 \\
$\Delta^{+}-\Delta^{++}$ & 
&0.54 & 0.08 & 2.08 &  &  & 0.36 \\
$\Delta^{0}-\Delta^{++}$ & $2.7\pm0.3$\protect\cite{PDB}
&3.21 & 2.20 & 6.10 &  &  & 2.54 \\
$\Delta^{-}-\Delta^{++}$ &  
& 8.04 & 6.34 & 11.38 &  &  & 6.55 \\
$\Sigma^{-}-\Sigma^{0}$ & $4.88\pm0.08$\protect\cite{PDB}
&7.09 & 5.16 & 6.07 & 2 & 4 & 5.24 \\
$\Sigma^{-}-\Sigma^{+}$ & $8.09\pm 0.16$\protect\cite{PDB}
&11.98 & 8.25 & 10.57 & 4 & 7 & 8.67 \\
$\Sigma^{*0}-\Sigma^{*+}$ & --4 to 4\protect\cite{PDB}
& 4.10 & 1.82 & 2.65 & 4 & 6 & 1.96 \\
$\Sigma^{*-}-\Sigma^{*0}$ & $2.0\pm2.4$\protect\cite{PDB}
& 6.34 & 3.85 & 4.40 & 3 & 4 & 3.69 \\
$\Xi^{-}-\Xi^{0}$ & $6.4\pm0.6$\protect\cite{PDB}
& 10.62 &  7.12 & 9.19 & 3 & 6 & 7.46 \\
$\Xi^{*-}-\Xi^{*0}$ & $3.2\pm0.6$\protect\cite{PDB}
& 5.87 &  3.68 & 3.58 & 3 & 3 & 3.58 \\
$\Sigma_{c}^{++}-\Sigma_{c}^{0}$ & $0.8\pm0.4$\protect\cite{PDB}
& 0.12 & 1.06 & 2.91 & --2 & 3 & 1.20 \\
$\Sigma_{c}^{+}-\Sigma_{c}^{0}$ & $1.4\pm0.6$\protect\cite{PDB}
& --0.96 & --0.55 & 0.55 & --2 & 1 & --0.36 \\
$\Xi_{c}^{0}-\Xi_{c}^{+}$ & $2.5\pm1.7\pm1.1$\protect\cite{CLEO}
& 4.67  & 2.58  &  2.90  & 2 & 0 & 2.83 \\
${\Xi'}_{c}^{0}-{\Xi'}_{c}^{+}$ & $1.7\pm4.6$\protect\cite{CLEO}
&1.04 & 0.47 &  --0.22  & 1  & 0  & 0.30 \\
$\Xi_{c}^{*0}-\Xi_{c}^{*+}$ &  $6.3\pm2.6$\protect\cite{CLEO}
& 0.40    & 0.44      & -0.85  & 1  & 0  & 0.43      \\
$\Sigma_{b}^{+}-\Sigma_{b}^{-}$ & &  --3.58 &  --3.45  &5.64  &  &  &   
--3.57
\\
$\Sigma_{b}^{0}-\Sigma_{b}^{-}$ & & --2.85 &  --1.99  &0.01  &  &  &   
--2.51 \\
$\Xi_{b}^{-}-\Xi_{b}^{0}$ & & 7.25  & 5.12  & 4.27 &   &   & 5.39 \\
${\Xi}_{cc}^{+}-{\Xi}_{cc}^{++}$ &-1.87 & -2.70   & -5.21   &    &    
&    & -2.96   \\
\hline
$\Sigma_{c}^{++}-\Sigma_{c}^{+}$ && 1.08 & 1.61 & 2.36 & 0 & 2 & 1.56  
\\
\hline
$\Sigma^{+}+\Sigma^{-}-2\Sigma^{0}$ &&
 2.20 & 2.07 & 1.57 &   &   & 1.81 \\
$\Sigma^{*+}+\Sigma^{*-}-2\Sigma^{*0}$ &&
 2.24 & 2.03 & 1.75 &   &   & 1.73 \\
$\Sigma_{c}^{++}+\Sigma_{c}^{0}-2\Sigma_{c}^{+}$ &&
 2.04 & 2.16 & 1.81 &   &   & 1.92 \\
\end{tabular}
\end{table}

\end{document}